Attention but not musical training affects auditory grouping

Sarah A. Sauvé & Marcus T. Pearce

Queen Mary University of London

Abstract

While musicians generally perform better than non-musicians in various auditory discrimination tasks, effects of specific instrumental training have received little attention. The effects of instrument-specific musical training on auditory grouping in the context of stream segregation are investigated here in three experiments. In Experiment 1a, participants listened to sequences of ABA_ tones and indicated when they heard a change in rhythm. This change is caused by the manipulation of the B tones' timbre and indexes a change in perception from integration to segregation, or vice versa. While it was expected that musicians would detect a change in rhythm earlier when their own instrument was involved, no such pattern was observed. In Experiment 1b, designed to control for potential expectation effects in Experiment 1a, participants heard sequences of static ABA_ tones and reported their initial perceptions, whether the sequence was integrated or segregated. Results show that participants tend to initially perceive these static sequences as segregated, and that perception is influenced by similarity between the timbres involved. Finally, in Experiment 2 violinists and flautists located mistuned notes in an interleaved melody paradigm containing a violin and a flute melody. Performance did not depend on the instrument the participant played but rather which melody their attention was directed to. Taken together, results from the three experiments suggest that the specific instrument one practices does not have an influence on auditory grouping, but attentional mechanisms are necessary for processing auditory scenes.

Attention but not musical training affects auditory grouping

Coined by Albert Bregman in his 1990 book, auditory scene analysis is the process by which we analyse the auditory world around us. The auditory system offers a 360° view of the world and provides information about objects that cannot be seen. Auditory streaming is the perceptual decomposition of sound input into its component sources, and has been the main conceptual approach to studying auditory scene analysis. It has been investigated in the context of sound attributes such as pitch (van Noorden, 1975), location (Jones & Macken, 1995), periodicity (Vliegen, Moore, & Oxenham, 1999) and timbre (Iverson, 1995) among others. Over the decades, a number of paradigms have been developed to explore and understand auditory streaming; two influential ones will be summarized here.

The first paradigm was pioneered by van Noorden (1975), in his doctoral research. It is a subtle but clever modification of the Miller & Heise (1950) paradigm: instead of alternating sounds in an ABAB pattern, van Noorden alternated sounds in an ABA- pattern, where '-' is a silence. This creates a triplet pattern when A and B are perceived as *integrated*, or coming from the same source, and an even pattern where the A stream is twice as fast as the B stream when A and B are perceived as *segregated*, or coming from two different sources. van Noorden (1975) explored the influence of pitch, tempo and loudness on this rhythmic perception and found that as pitch, tempo and loudness difference increased, perception tended towards segregation. In other words, the more different A and B are, the more likely they are to be perceived as coming from different sources. van Noorden further defined segregation parameters with fission and temporal coherence boundaries. While the fission boundary defines the difference below which integration is inevitable (the pitch, tempo or loudness are too similar or slow to lead to segregation), the temporal coherence boundary defines the difference above which segregation is inevitable (the pitch, tempo or loudness are too dissimilar or fast to allow for integration). Between these two boundaries, perception is bi-stable, meaning that either integration or segregation are possible and depend on other factors (Denham & Winkler, 2006). This paradigm has been used to help researchers understand how various sound attributes contribute to our perception of the world around us (Rose & Moore, 2000; Singh & Bregman, 1997; Vliegen, Moore, & Oxenham, 1999). For example, that streaming occurs before temporal integration of the auditory scene (Yabe et al., 2001) and that both spectral content and location interact in the processing of ambiguous auditory scenes (Shinn-Cunningham, Lee, & Oxenham, 2007).

Another prominent paradigm, using slightly more complex stimuli, was introduced by Dowling (Dowling, 1973) and was named the interleaved melody paradigm. Here, the notes of two melodies are presented in an alternation, such that melody 'ABCDEF' and melody 'abcdef' become 'AaBbCcDdEeFf'. Dowling found that as pitch overlap decreased, participants were more easily able to detect, or segregate, each individual melody. Trained musicians could tolerate more pitch overlap than non-musicians. The concept can similarly be applied with many other parameters including loudness and timbre (Hartmann & Johnson, 1991), where it is easier to track a melody if the two interleaved melodies are of different loudness, or played by different instruments.

Timbre is a complex auditory parameter and timbral perception has been investigated in detail using both synthesized tones and real instrumental sounds (Alluri & Toiviainen, 2010; Caclin, McAdams, Smith, & Winsberg, 2005; McAdams, Winsberg, Donnadieu, De Soete, & Krimphoff, 1995). The most common method of investigating timbre has been multidimensional scaling, or MDS. Based on dissimilarity ratings between pairs of timbres, sounds are mapped into a multi-dimensional space representing perceptual distance. In research to date, three dimensions seems to provide an optimal representation of perceptual timbre space; though the first two are fairly stable across experiments, the third is less well established. The first two represent log rise time (the attack), and spectral centroid while the third dimension that emerges is usually a spectro-temporal feature such as spectral flux or spectral irregularity. One of the biggest issues with this research however is that in most cases the rated sounds are synthesized (though see Kendall & Carterette, 1991, and Lakatos, 2000, for examples of MDS using natural stimuli). Besides this, our perceptual system is not used to hearing synthetic sounds such as these and may process them differently than natural sounds (Gillard, J. & Schutz, M., 2012). Therefore, it is important to complement studies using controlled synthesized tones with investigations using natural sounds.

The role of musical training has been extensively studied in the context of auditory skills, including auditory streaming (François, Jaillet, Takerkart, & Schön, 2014; Zendel & Alain, 2009). As a result of training, musicians are more sensitive to changes in auditory stimuli based on pitch, time and loudness for example (Marozeau, Innes-Brown, & Blamey, 2013; Marozeau, Innes-Brown, Grayden, Burkitt, & Blamey, 2010), with discrimination thresholds being lower in musicians than in non-musicians. One problem with treating musicians as a single category is that differences between instrumentalists may be missed (Tervaniemi, 2009). Pantev and colleagues (Pantev, Roberts, Schulz, Engelien, & Ross, 2001) found that certain instrumentalists were more sensitive to the timbre of their own instrument than to others, as measured by auditory evoked fields (AEF). Violinists and trumpet players were presented with trumpet, violin and sine tones while MEG was recorded. Both instrumentalists presented stronger AEFs for complex over sine tones, and stronger AEFs still for their own instrument. In a similar study (Shahin, Roberts, Chau, Trainor, & Miller, 2008), professional violinists and amateur pianists as well as young piano students and young non-musicians were presented with piano, violin and sine tones while reading or watching a movie and EEG was recorded. Gamma band activity (GBA) was more robust in professional musicians for their own instruments and young musicians showed more robust GBA to piano tones after their one year of musical training. Furthermore, Drost, Rieger, & Prinz, (2007) found that pianists and guitarists' performance on a performance task was negatively affected by auditory interference, but only if it was their own instrument. Taking a step further and using more ecological stimuli, Margulis, Mlsna, Uppunda, Parrish, & Wong, (2009) explored neural expertise networks in violinists and flautists as they listened to excerpts from partitas for violin and flute by J. S. Bach. Increased sensitivity to syntax, timbre and sound-motor interactions were seen for musicians when listening to their own instrument.

More recently, pianists, violinists and non-musicians listened to music during fMRI scanning (Burunat et al., 2015). The authors investigated the effects of musical training on callosal anatomy and interhermispheric functional symmetry and found that symmetry was increased in musicians, and particularly in pianists, in visual and motor networks. They concluded that motor training, including differences between instrumentalists, affects music perception as well as production. Other research has investigated differences between types of musical training. For example, one study used EEG to show that conductors have improved spatial perception, when compared to non-musicians and pianists (Nager, Kohlmetz, Altenmuller, Rodriguez-Fornells, & Münte, 2003). Another line of research investigates pianists' formation of action-effect mappings due to the design of their instrument (Baumann et al., 2007; Drost, Rieger, Brass, Gunter, & Prinz, 2005; Repp & Knoblich, 2009; Stewart, Verdonschot, Nasralla, & Lanipekun, 2013).

However, such specific effects of instrumental training have not yet been observed in auditory streaming, where an effect would be seen by a change in streaming threshold. We hypothesise that with increased sensitivity to a particular timbre, it would take less time to detect two separate auditory objects when one's own instrument is one of these objects. This is the basis of the first experiment reported in this paper.

The objective of this research is to test the hypothesised increased sensitivity in streaming the instrument(s) which a musician plays. Three experiments will be presented. The first is a classic ABA_ streaming paradigm, the second is a control study that examines the effects of prior expectation and the third is an interleaved melody paradigm, designed to corroborate findings in Experiments 1a and 1b using more musically realistic stimuli.

### Experiment 1a

The ABA_ paradigm (van Noorden, 1975) is used here and timbre is manipulated instead of pitch. While the timbre of a *standard sequence* remains static throughout a given trial, a *target sequence* morphs from one timbre to another, creating a qualitative change from a galloping ABA_ rhythm to the perception of two simultaneous, isochronous A_A_A and B___B___B patterns as the standard and target sequences' timbres become more and more different, or vice versa as the timbres become more similar. The point of change in rhythmic perception reflects the detection of a new sound object, or, in the other direction, the merging together of two sound objects. The sound objects (standard and target streams) are defined solely by their timbre, as pitch, length and loudness are controlled. Based on previous work (Sauvé, Stewart, & Pearce, 2014), detection of a sound object defined by one's own instrumental timbre is predicted to occur sooner than for other instrumental timbres, when the participants' instrument is the target (i.e. it is 'new to the mix' and captures attention) and later than for other instrumental timbres when the instrumentalists' timbre is the standard (i.e. it already holds attention and delays perception of the arrival of a new sound object). This previous study compared seven different instrumental timbres in the same ABA_ paradigm, while additionally exploring the effect of attention on streaming by manipulating participants' attentional focus. Results guided the design of the current study by providing target effect sizes, refining the test timbres and allowing the elimination of the attention manipulation, as it was confirmed to have a significant impact on the perception of auditory streams.

**Method**

**Participants.** Participants were 20 musicians (13 females, average age 34.45; SD = 7.59; range 21-69) recruited from universities and the community. Their average Gold-MSI score (Müllensiefen, Gingras, Musil, & Stewart, 2014) for the musical training subscale was 40.15 (SD = 4.23); 5 were violinists, 6 were cellists, 5 were trumpet players and 4 were trombone players.

**Stimuli.** All four timbral sounds (violin, trumpet, trombone, cello) were chosen from the MUMS library (Opolko & Wapnick, 2006) with pitches spanning an octave (all 12 pitches between A220 to G#415.30). The files were adjusted to equal perceptual length of 100ms and equal loudness, based on the softest sound. A 10ms fade out was applied to each timbral sound. All editing was done in Audacity and the final product was exported as a CD quality wav file (44,100 Hz, 16 bit). See Appendix A for full details.

Using a metronome in Max/MSP, the *standard sequence* was presented by playing a selected timbre with an inter-onset interval of 220ms. The *target sequence* was presented using another metronome at a rate of onset of 440ms, beginning 110ms after the standard sequence to create the well-known galloping ABA_ pattern (van Noorden, 1975). The target sequence was a series of 100ms sound files representing a 30s morph between the standard timbre and the target timbre, achieved using a slightly modified Max/MSP patch entitled 'convolution-workshop'. This patch is distributed by Cycling '74 with Max/MSP. The target sequence morphed from standard to target timbre in the ascending condition, creating a galloping to even rhythm change, and from target to standard timbre in the descending condition, creating an even to galloping rhythm change (see Figure 1). Each trial ended when the participant indicated a change in perception or after 30s if participants did not reach a change in perception.

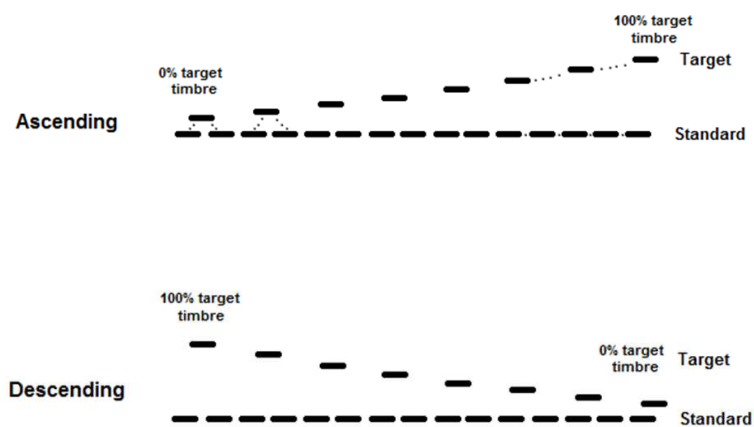

**Figure 1**. Illustration of ABA_ paradigm, ascending and descending, modifying timbre only.

**Procedure.** The experiment was coded and run in Max/MSP, with output presented through headphones and input taken from mouse clicks. Participants were first presented with a practice block with instructions and an opportunity to listen to each timbre and rhythm separately. Up to four practice trials were included in the block and questions were welcomed. Participants then began the first of two experimental blocks.

For each trial, participants indicated by clicking a button on the screen at which point the galloping sequence became perceived as two separate streams of standard and target tones, or the opposite for descending presentation. This point was recorded as the percent of time passed in the trial, which equates to the percent of morphing at that time. Each trial lasted a maximum of 30s, at which point the trial ended automatically and a value of '-1' was recorded, indicating that the participant had not reached a change in perception on that trial. Trials were presented in two blocks, and participants were instructed to indicate a change in rhythm as soon as it was perceived for the ascending block and to hold on to the original rhythm as long as possible for the descending block. Together, this gives two measures of the fission boundary (van Noorden, 1975). The fission boundary was measured instead of the temporal coherence boundary due to its higher sensitivity for detecting timbral effects in perception, and due to confirmation that the fission and temporal coherence boundaries are separate phenomena that can be manipulated by instruction (Sauvé et al., 2014). For every block, every timbre modulated to every other timbre once for a total of 12 trials (4 timbres each modulating to the 3 other timbres), each separated by 4s and each at a different pitch, to reduce trial to trial expectancy and habituation. Participants were randomly assigned to one of two different orders to control for any order effects.

Once both blocks were completed, participants filled out the musical training sub-scale of the Goldsmiths Musical Sophistication Index (Müllensiefen et al., 2014).

**Analysis.** Effect sizes and confidence intervals were used in the analysis of Experiments 1a and 1b, in addition to traditional methods. These methods are based on Cumming (2012; 2013), who advocates wider use of effect sizes and confidence intervals in the research community to increase integrity, accuracy and the use of replication. According to Cumming, the low occurrence of null results in the literature and a pressure towards new studies and away from replication translates into misrepresentation and inhibition of scientific knowledge. Cumming advocates the use of effect sizes, confidence intervals, and meta-analysis in place of null hypothesis significance testing (NHST). This method is preferred because confidence intervals give more information both about the current effect size, and about potential future replications by offering a range of potential values for a measure, rather than one indicator of significance or non-significance. For more information about effect size and confidence interval methods, see Cummings' book, *The New Statistics* (2012) or the corresponding article for a shorter summary (2013).

**Results**

Percentage of time passed (degree of morphing) is the dependent variable analysed; for descending trials the percentage was subtracted from 100 so that ascending and descending conditions can be compared directly. A low percentage indicates early streaming in the ascending condition and late integration in the descending condition while a high percentage indicates late streaming in the ascending condition and early integration in the descending condition. Furthermore, trials in the ascending condition where the percentage exceeded 100 were replaced with 100 and trials in the descending condition where the percentage was negative were replaced with 0. These are all cases where the participant listened to the trial for more than 30 seconds and still did not hear a change in rhythm. Five participants' data were removed because they did not hear a change in rhythm in more than half of the trials, in either or both blocks (two violinists, two cellists and a trumpet player). The difference between mean percentage for ascending and descending conditions was 1.2, 95% CI [-4.4, 6.8]. As the CIs include zero, the difference was not significant. However, mean percentage of time passed was significantly higher for the first block of trials than the second, with a difference of 10.6 [2.6, 18.6] for the ascending and 10.7 [3.1, 18.1] for the descending conditions. As both CIs do not include zero, the difference is significant.

Effects of specific instrumental training were investigated next. Data were grouped by instrumentalist and then sub-grouped by standard timbre. For violinists, mean percent time passed when violin was the standard timbre was 56.5 [50.7, 62.3], mean percent for cello was 59.8 [53.5, 66.1], mean percent for trumpet was 65.8 [51.8, 79.8] and mean percent for trombone was 64.6 [50.8, 78.4]. See Table 1 for details of all instrumentalists. Data were then sub-grouped by target timbre. When violin was the target timbre, mean percent for violinists was 62.1 [50.0, 74.2], mean percent for cellists was 48.2 [36.9, 59.5], mean percent for trumpeters was 54.5 [44.5, 64.5] and mean percent for trombonists was 48.4 [38.1, 58.7]. See Table 1 for details of all target timbres. Figure 2 displays results graphically.

Thresholds for an instrumentalists' own timbre were hypothesised to be lower when their own instrument was the target and higher when it was the standard. However, interpreting the CIs above does not reveal any reliable pattern of results. If more than half the margins of error (MOE), which is one half of the CI, overlap when comparing between subject groups, the difference is not considered significant. While two comparisons attain significance (trombone players have a lower threshold than trumpet players for the trombone sound as standard, and trombone players have a lower threshold than trumpet players for the trumpet sound as target), this is not enough to establish a pattern. Comparison of confidence intervals cannot be done so easily for within-subject measures, therefore a mixed effects model was applied, where instrument played and standard, or target timbre, predicted threshold. The instrument played had no effect on perceptual threshold, $\chi^2 (3) = 3.83$ and $\chi^2 (3) = 3.82$, $p = .28$ for standard and target models respectively.

**Table 1**. Mean percent of trial duration by standard and target timbre, and by instrumentalist, with 95% confidence interval margins of error (MOE).

|  |  | Mean Duration ± MOE | | | |
|---|---|---|---|---|---|
|  |  | **Violin** | **Cello** | **Trumpet** | **Trombone** |
| **Standard Timbre** | Violinist | 56.5 ± 5.8 | 59.8 ± 6.3 | 65.8 ± 14.0 | 64.6 ± 13.8 |
|  | Cellist | 50.7 ± 12.0 | 46.3 ± 12.1 | 55.7 ± 11.9 | 47.8 ± 11.0 |
|  | Trumpeter | 53.7 ± 11.8 | 50.8 ± 6.6 | 50.0 ± 10.3 | 59.6 ± 12.1 |
|  | Trombonist | 49.4 ± 13.9 | 57.0 ± 9.1 | 48.0 ± 9.9 | 39.5 ± 9.5 |
| **Target Timbre** | Violinist | 62.1 ± 12.1 | 62.1 ± 13.4 | 62.9 ± 9.1 | 62.0 ± 9.8 |
|  | Cellist | 48.2 ± 11.3 | 51.3 ± 11.4 | 46.8 ± 13.9 | 53.2 ± 11.6 |
|  | Trumpeter | 54.5 ± 10.0 | 47.3 ± 9.0 | 67.1 ± 11.4 | 49.2 ± 9.8 |
|  | Trombonist | 48.4 ± 10.3 | 45.3 ± 12.3 | 45.4 ± 11.4 | 53.4 ± 7.4 |

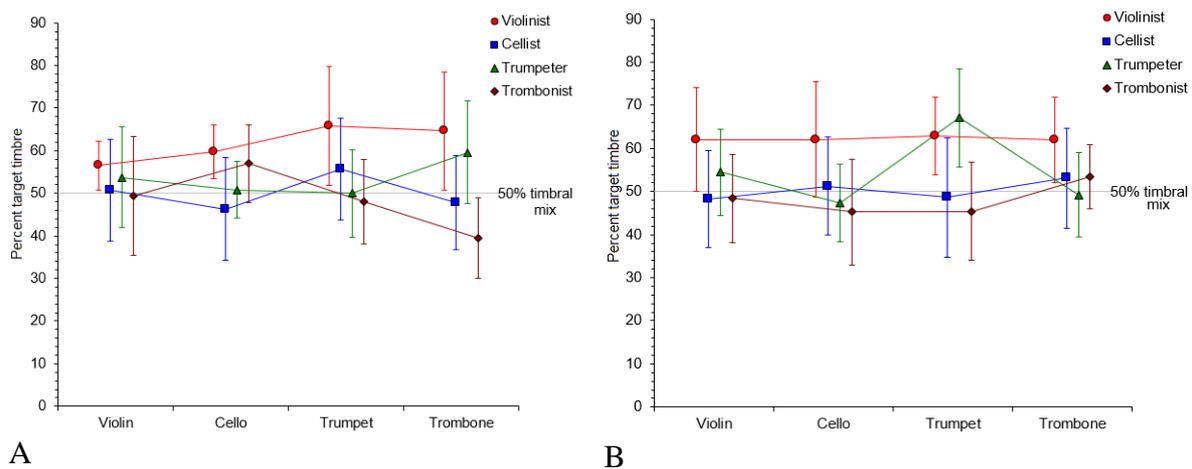

**Figure 2**. Percent target timbre contained in the morphing stream at the point of a change in percept as a function of instrumentalist, and standard (A) and target (B) timbres. Error bars represent 95% CIs.

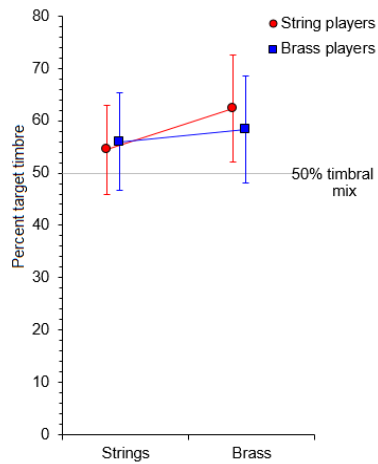

**Figure 3**. Percent target timbre of the morphed stream at the point of a change in percept for brass and string instrumental family groups. Error bars represent 95% CIs.

Effects of instrumental family were also investigated. Performance by instrumental group was analysed for string pair and brass pair trials (i.e. where the standard and target timbres were both string or both brass instruments). String players performed with a mean percentage of 54.5 [46.0, 63.0] on string pairs and 62.4 [52.2, 72.6] on brass pairs. Brass performed with a mean percentage of 56.0 [46.7, 65.3] on string pairs and 58.4 [48.2, 78.6] on brass pairs (see Figure 3). Interpreting the CIs indicates that there was no difference between string players and brass players; however, a mixed effects model to investigate within group differences found that instrument played had an effect on threshold, $\chi^2 (1) = 3.54$, $p = .05$, where string players had a lower discrimination threshold for string instruments than for brass instruments.

Trials where participants did not hear a change in rhythm were examined separately. Most participants only had a few trials where this happened, if at all. As noted above, for five, this case was more prominent and their data were removed (it is interesting to note that the mean age for these five participants is 51.8 (SD = 12.7) and every participant was at or above the average age for all participants). Every type of instrumentalist was represented in this group of trials; all for the ascending block and all but cellists for the descending blocks. The frequency of each of the standard and target timbres was different within each direction by timbre type condition (i.e. the number of times a trial had cello as the standard or target timbre, versus the other instruments, in the ascending or descending block), but no single timbre was consistently more or less represented. When looking at pairs of timbres, the cello-trumpet and trombone-trumpet pairs were most commonly still perceived as an even percept by the end of a trial in the descending condition and the violin-trombone pair was the most commonly still perceived as a galloping percept by the end of a trial in the ascending condition.

**Discussion**

This experiment was designed to corroborate neuroscientific measures showing that instrumentalists are more sensitive to their own instrument's timbre than to others (Pantev et al., 2001). Accordingly, in the ABA_ paradigm, we hypothesised a lower timbre discrimination threshold for instrumentalists hearing their own instrument when their instrument is the target timbre, and a higher discrimination threshold when their instrument is the standard timbre.

Results show no reliable effect of instrument played on the perception of timbral stream segregation when looking at individual target instruments. Though thresholds for an instrumentalists' timbre were slightly lower than for other timbres when looking at standards, contrary to the hypothesis, none of these differences were significant. Similarly for target timbres, no threshold differences were significant, though the largest effect was seen in trumpet players, where the threshold when trumpet was the target was higher than for other instruments. There was a small effect of instrument played when comparing performance on instrumental families: string players detected the difference between two brass instruments later than for two string instruments. They were not better than brass players at detecting the difference between two string instruments, nor did brass players show an advantage for brass instruments. Thresholds for string instruments were overall lower than for brass instruments. Perhaps the two string instruments were more different than the two brass instruments, thus making them overall easier to distinguish (this is supported by timbre dissimilarity ratings collected in Experiment 1b). The effect of order is unexpected and could be the result of a familiarization with the task that led to greater sensitivity in the second block.

How can such results be explained when the literature reviewed, particularly Pantev's work (2001), suggests an effect of instrumental training on perception? Let us first place the question in a more generalized context. Imagine a trained musician is listening to an orchestral work. Just like most listeners, they clearly hear the melody. What if they were asked to listen to the bass line? Or another instrument? If instrumentalists are more sensitive to their own instrument's timbre, then it would be expected that they could more easily and more accurately pick out (and perhaps transcribe, for potential experimental purposes) their own instrument than any other. However, according to the present results, they could also pick any instrument out of the auditory scene and transcribe it just as well. This would suggest that ability to pick out and transcribe a particular line in a polyphonic work is not related to the instrument one plays, but rather to general musical training, and to where their attention is directed. It would be interesting to conduct a transcription experiment along these lines in future research. However, a reasonable explanation of the present results is that listeners simply heard what they paid attention to, though it is only a proposition here and cannot be supported or countered with the current data. The possibility of attention directing perception will be further explored in Experiment 1b and Experiment 2.

One of the basic claims of auditory streaming is that coherence is the default percept (Bregman, 1978; Bregman, 1990; Rogers & Bregman, 1998). However, if this were the case, then initial segregation in the descending condition of this experiment would not be possible. The fact that participants were told what they would be hearing (even to galloping for descending blocks and galloping to even for ascending blocks) could have influenced their perception of the stimuli by setting up a specific expectation. Therefore, an experiment to control for this was designed and is reported next.

## Experiment 1b

This experiment was designed to control for the possible expectation effect of the instructions given in Experiment 1a. Participants were presented with 10s of ABA_ pattern where the timbres are unchanging and maximally different (the same as the beginning of a descending block trial in Experiment 1a) and were asked to report whether they heard an even or a galloping pattern. If participants tend to hear these stimuli as even, then there is cause to revisit the default coherence concept; alternatively, if participants tended to hear the stimuli as galloping, then the instructions given in Experiment 1a likely set up an expectation which strongly influenced perception, enough to hear an even pattern at first hearing. Participants were also asked to indicate which of the two timbres was most salient. If the standard timbre (the faster stream) is chosen most often then timing tends to attract attention more than timbre; if the standard and target timbres are chosen approximately equally often, then it is the timbre itself that is most salient in capturing focus.

**Method**

**Participants.** Data was collected in two groups: first, undergraduate and graduate musicians and, second, individuals with various other backgrounds recruited from universities in London and the community. The first group of participants were the same 20 participants as in Experiment 1a (the same five participants' data was excluded here); they completed both paradigms. The second group was tested separately and included a wider range of backgrounds to control for effects of musical training in the first group. This second group consisted of 20 individuals (7 males, mean age 22.5 years; SD = 4.33; range = 18-32; mean Gold-MSI score = 23.3, SD = 11.9, range = 7-46) recruited through volunteer email lists, credit scheme and acquaintances. Participants in the first group were entered in a draw for an Amazon voucher while participants in the second group were either entered in a draw for an Amazon voucher or given course credit as part of a university credit scheme.

**Stimuli.** The stimuli were the same as Experiment 1a, except that there were seven timbres (piano, violin, cello, trumpet, trombone, clarinet, bassoon) and there was no morphing. One timbre was presented at 220ms and the other at 440s with a 110ms offset and the total length of one trial was 10s.

**Procedure.** This paradigm was also presented in Max/MSP. After reading the information sheet and giving written consent, instructions were presented on the screen along with examples of the even and galloping patterns, each accompanied by an illustration to help clearly distinguish the two rhythms. Five practice trials were provided and were compulsory, giving a chance for questions and clarification before beginning the data collection.

When ready to begin, for each trial participants indicated as they were listening which percept they heard first using the keyboard, pressing 'H' (horse) for the galloping pattern and 'M' (morse) for the even pattern (terminology from Thompson, Carlyon, & Cusack, 2011). At the end of the trial, they clicked on the timbre that was most salient to them (the appropriate two were displayed at each trial). Every possible timbre pair was explored, for a total of 21 trials.

Participants then completed the musical training sub-scale of the Gold-MSI (Müllensiefen et al., 2014).

**Timbre dissimilarity ratings.** Timbre dissimilarity ratings were collected separately using Max/MSP. 15 listeners of varying backgrounds, none of which participated in the reported experiments, rated the similarity of pairs of timbres on a 7-point Likert scale where 1 was the least similar timbre pair and 7 was the most similar timbre pair, with other pairs rated between these numbers. The participants could listen to seven musical tones at any time. These were the same as in Sauve et al. (2014) (piano, violin, cello, trumpet, trombone, clarinet, bassoon). Participants clicked a button to begin a trial: two timbres were presented for comparison and participants rated the similarity between the sounds. There was no time limit and participants submitted each rating on their own time, completing the trial. Pairs of timbres were presented randomly. Results are shown in Figure 4.

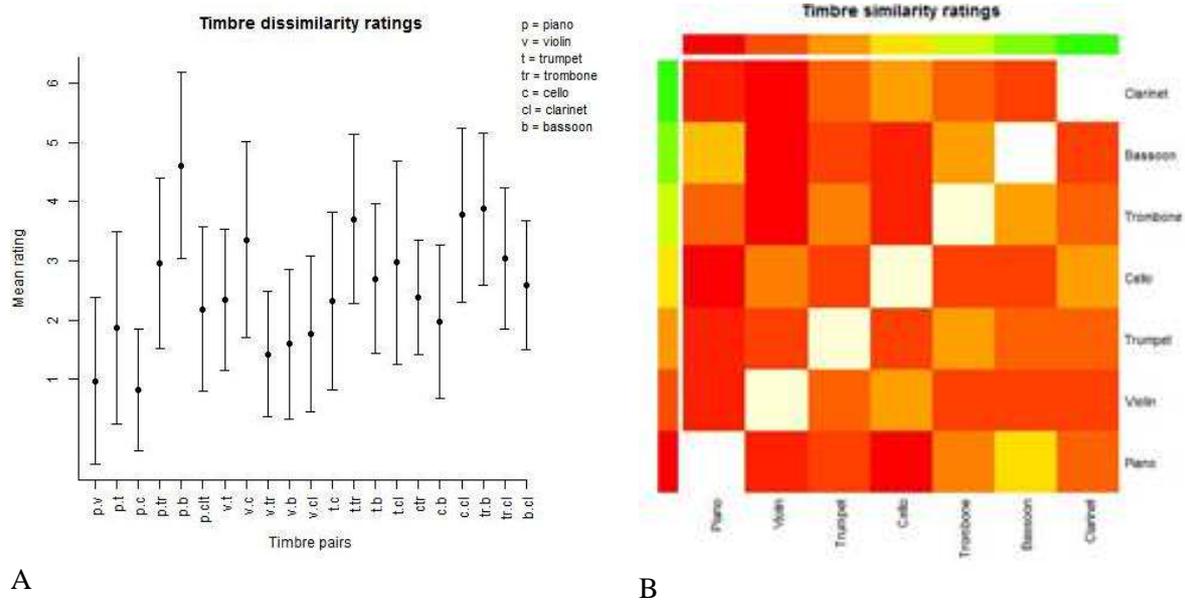

**Figure 4**. A. Timbre dissimilarity ratings (1-7 Likert scale; 1 is very dissimilar, 7 is very similar). When the initial percept is even, timbres are less similar (2.30 [2.22, 2.38]) and when the initial percept is galloping, timbres are more similar (2.90 [2.80, 3.00]). B. Timbre dissimilarity ratings presented in a heat map, where red is most dissimilar and green is most similar.

**Results**

A comparison of the two groups revealed no significant difference between the initial percept for musicians and for non-musicians; difference in proportions were .03 [-.04, .10]. Therefore the remaining analysis was performed on aggregated data.

The mean of the initial percept, where even was coded as 0 and galloping was coded as 1, was .35 [.32, .39]. Interpreting the CIs in Figure 5 indicates that this is significantly different from chance (.5). Because the mean of initial percept is closer to zero than it is to one, the initial percept is dominantly even.

A 'matching' variable was created, where if the timbre identified as salient matched the standard timbre, a value of 1 was assigned and if it did not, a value of 0 was assigned. The mean of the matching variable was .69 [.65, .72]. Once again, interpreting the CIs in Figure 5 indicates that this is significantly different from chance (.5), confirmed by an exact binomial test, $p < .01$. Therefore, the most salient timbre is most often the standard timbre.

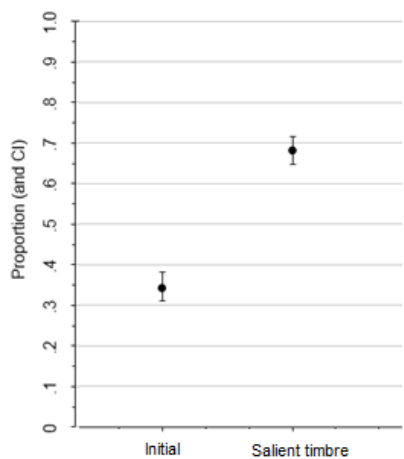

**Figure 5**. Mean of initial (left) and matching (right) variables, both significantly different from chance.

The influence of timbre was investigated using timbral dissimilarity ratings to assess whether more similar timbre pairs would encourage integration while less similar pairs would encourage segregation. This pattern was observed in the data. The average dissimilarity rating over all trials where segregation was the initial percept was lower, 2.30 [2.22, 2.38], than when integration was the initial percept, 2.90 [2.80, 3.00], confirmed by $t(499) = -8.11$, $p < .01$.

**Discussion**

This experiment was designed to investigate whether the instructions in Experiment 1a enabled the possibility of initial segregation in the descending blocks by setting up the expectation for segregation, as according to streaming theory, integration is always the default percept until enough evidence is gathered for the existence of two separate streams (Bregman, 1990).

Results indicate that the even percept is the most common initial percept, which is contrary to the streaming theory discussed above. However, this experiment does not rule out the possibility that the build-up of evidence for two streams simply happened very quickly. A reliable neural streaming marker is needed to investigate this question at the millisecond level. While some such markers have been suggested (Alain, Arnott, & Picton, 2001; Fujioka,

Trainor, & Ross, 2008; Sussman, Ritter, & Vaughan, 1999), none of them constitute direct measures of streaming.

Furthermore, the initial percept depended on how similar pairs of timbres were. It would presumably take longer for the brain to find evidence for two streams if the sources were more similar, and less time if they were less similar. A similar pattern for pitch was found by Deike et al. (2012), where participants were presented with ABAB sequences and asked to indicate as quickly as possible whether they heard one or two streams. The separation between A and B tones varied from 2 to 14 semitones. Results showed that the larger the pitch separation between A and B tones, the more likely participants were to hear the sequence as segregated in the first place. Predictability was also found to influence degree of segregation (Bendixen, Denham, & Winkler, 2014): when degree of predictability between two interleaved sequences was high, an integrated percept was supported, while when the predictability within each interleaved sequence alone was high, a predominantly segregated percept was induced. This is contrary to the integration-by-default concept proposed by Bregman (1990). However, auditory scene analysis is complex and we have not addressed the role of context, which has been shown to speed or slow the buildup of evidence for perceptual segregation (Sussman-Fort & Sussman, 2014).

Attentional mechanisms were probed by asking participants which timbre was most salient. Results show that the standard timbre was most often the most salient timbre. In feedback, some participants described it as more driving and therefore more attention-drawing. This suggests that rhythm is a more salient feature than timbre, adding interesting evidence to discussions about the relative salience of different features in the perception of polyphonic music (Esber & Haselgrove, 2011; Prince, Thompson, & Schmuckler, 2009; Uhlig, Fairhurst, & Keller, 2013).

In terms of the influence of instructions in Experiment 1a, it seems that they did influence participants' perception; otherwise, initial segregation on trials with similar pairs of timbres would not be possible. It is already known that attention influences perception in this paradigm (Sauvé et al., 2014), and this experiment suggests that prior expectation about the number of streams also has an impact.

**Experiment 2**

Experiment 1a aimed to behaviourally test the hypothesis that instrumentalists are more sensitive to their own instrument's timbre than to others. Experiment 1b was designed to control for the effect of expectation. However, neither of these paradigms are particularly ecologically valid; the ABA_ pattern is especially synthetic and though the sounds are recorded and not synthesized, the way they are combined is not reminiscent of actual music. Experiment 2 was designed with the same goal as Experiment 1a and to allow results to be extended towards more ecological musical listening. The interleaved melody paradigm introduced by Dowling (1973) was selected to achieve this goal. The task was to detect one or multiple mistunings, as intonation is a developed skill in many instrumentalists. In the original interleaved melody paradigm, Dowling asked participants to identify the melodies being played and found that this was more likely to occur when pitch overlap between the two melodies was minimal. With increased sensitivity, more pitch overlap is possible; for example, musicians are able to identify melodies with more pitch overlap than non-musicians. Similarly, it is hypothesized that instrumentalists should identify mistunings more accurately for their own instrument overall, and with more pitch overlap as well.

**Method**

**Participants.** Participants were 15 musicians, 8 flautists and 7 violinists, recruited from music schools and conservatoires in London and in Canada. If desired, they were entered in a draw for one of two Amazon vouchers.

**Stimuli.** Melodies were two excerpts from compositions by J. S. Bach: BWV 772-786 Invention 1, mm13 and BWV 772-786, Invention 9, mm14-15.1 (only the first beat of mm15). They are in different meters (4/4 and 3/4 respectively) and different keys (A minor and F minor respectively), but have similar ranges (perfect $12^{th}$ - octave + perfect $5^{th}$ - and diminished $12^{th}$ - octave + diminished $5^{th}$ - respectively) and similar median pitches (C#4 and B4 respectively). The 4/4 melody was played on a violin and the 3/4 melody on a flute.

A violinist and a flautist were recorded using a Shure SM57 microphone, recorded into Logic and exported as CD quality audio files. These original recordings were verified by a separate violinist and flautist for good tuning and corrections to tuning were made using Melodyne Editor by Celemony. Melodies were recorded at notated pitch and for every necessary transposition to create each overlap condition as tuning in a solo instrument changes slightly as a function of key, especially in Baroque music (just intonation).

Using Melodyne Editor, 50 cent sharp mistunings were inserted. Each trial contained either zero, one or two mistunings. The location of each mistuning is presented in Table 2. Though it is recognized that sharp or flat tuning may be perceived differently and depends on the context (Fujioka, Trainor, Ross, Kakigi, & Pantev, 2005), only one direction was used here for simplicity. The tempo and note length of the melodies were quantized, and the melodies interleaved, so that the onset of the first note of the second melody fell exactly between the onsets of the first and second notes of the first melody, the second between the second and the third, and so on.

Table 2. Experimental design: details of metrical and instrument location of mistunings (where there are two mistunings, these are separated by a backslash), the higher melody, where attention was directed and the amount of pitch overlap between the mean pitch of the two melodies for each trial, including practice and control trials.

| Trial | Location (metrical) | Location (instrument) | Top melody | Attentional focus | Pitch overlap |
|---|---|---|---|---|---|
| Practice 1 | 4.1 | Violin | Violin | Violin | 5$^{th}$ |
| Practice 2 | 1.3 / 2.4 | Flute / Flute | Flute | Flute | 5$^{th}$ |
| 1 | 1.3 / 3.3 | Flute / Violin | Flute | Flute | 2$^{nd}$ |
| 2 | 2.4 / 4.2 | Violin / Violin | Flute | Violin | 2$^{nd}$ |
| 3 | None | None | Violin | Flute | 2$^{nd}$ |
| 4 | 3.1 | Violin | Violin | Flute | 2$^{nd}$ |
| 5 | 2.2 | Flute | Flute | Flute | 3$^{rd}$ |
| 6 | 1.4 / 2.1 | Flute / Violin | Flute | Both | 3$^{rd}$ |
| 7 | 1.2 / 3.4 | Violin / Flute | Violin | Flute | 3$^{rd}$ |
| 8 | 3.2 / 4.3 | Flute / Flute | Violin | Violin | 3$^{rd}$ |
| 9 | None | None | Flute | Violin | 5$^{th}$ |
| 10 | 2.4 / 4.1 | Flute / Flute | Flute | Flute | 5$^{th}$ |
| 11 | 2.3 / 4.1 | Violin / Violin | Violin | Violin | 5$^{th}$ |
| 12 | 3.2 / 4.2 | Violin / Flute | Violin | Both | 5$^{th}$ |
| Control 1 | 2.3 | Violin | - | - | - |
| Control 2 | 3.1 | Flute | - | - | - |

Twelve experimental trials were created, along with two practice trials and two control trials. Five variables were manipulated: metrical mistuning location, instrumental mistuning location, top melody, attentional focus and pitch overlap. The mistunings were either on strong or weak beats; location is indicated by beat (first number) and subdivision (second number) i.e. 4.2 = beat 4, second subdivision (sixteenth note). The mistunings were either in the violin or the flute melody, the top (also the first tone heard) melody was either the violin or the flute melody and the participants' focus was directed at either the violin melody, the flute melody, or both. Pitch overlap was either a 2$^{nd}$, a 3$^{rd}$ or a 5$^{th}$, where the distance between the central (in terms of range) pitches of each melody matched these intervals. The instrumental mistuning location, top melody and attentional focus were manipulated so that they sometimes match and sometimes do not (i.e. the mistuning may not be in the same melody to which the participant is asked to attend). This was intended to assess whether a mistuning in the non-attended melody influences identification of mistunings in the attended melody.

The control trials were single melodies, designed to ensure that participants were able to detect mistunings in a simpler listening situation. In a pilot study, a 50 cent mistuning in a single melody was always detected.

**Procedure.** This experiment was carried out online, using the survey tool Qualtrics. Once presented with the information sheet and detailed instructions, participants could give informed consent. The two original melodies (with no mistunings) were both presented for participants via SoundCloud to familiarize themselves with the tunes, and in every subsequent trial in case participants wanted to refresh their memory. Each page of the survey contained the two original melodies, the current trial (also via SoundCloud) and a click track. Participants clicked on the beats where they heard a mistuning; this was set up using Qualtrics' hot spot tool. There was one click track for trials where focus was on one instrument and two, stacked vertically and labelled with the corresponding instrument, when participants were instructed to listen to both (see Figure 6). The word 'none' under the click track was also a selection option if participants detected no mistuning.

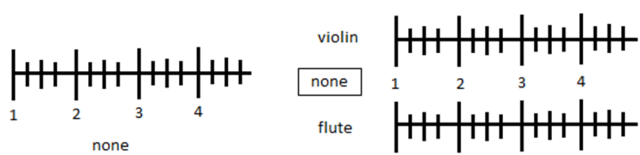

**Figure 6.** Single (left) and double (right) click tracks presented to participants alongside the relevant audio files.

Participants started with two practice trials, always in the same order. Then, trial 11 was always presented first because it was one of the trials with the least amount of overlap (and, therefore, presumably easier) and all other trials followed in random presentation. Finally, the two control trials were presented, always in the same order.

Participants finally selected their primary instrument, either violin or flute, and had the option to submit their email address for the Amazon voucher draw.

**Results**

Initial inspection of the data showed a high rate of false alarms. Participants were first screened by performance on the control trials; only participants who had correctly identified the mistunings in both control trials, without false alarms, were included in analysis. This left 12 participants; 6 violinists and 6 flautists.

A mixed effects binomial logistic regression was performed, with musicianship (violinist or flautist), metrical mistuning location, instrumental mistuning location, top melody, attentional focus and pitch overlap as predictors for accuracy and random intercepts on participants. Accuracy was simply defined by the number of correctly identified mistunings. Only attentional focus and top melody were strong significant predictors, $z(1) = -4.85$ and $z(1) = 3.65$ respectively, both $p < 0.01$ while instrumental mistuning location was moderately significant, $z(1) = 2.15$, $p = .03$. There were no significant interactions (see Table 3 for details).

Accuracy when attention was directed to the violin line was highest, at .28 [.22, .35], to the flute line was .25 [.20, .32] and to both was lowest, at .10 [.06, .17]. Accuracy when the violin line was on top was lower than when the flute was on top, at .17 [.13, .22] and .29 [.24, .34] respectively. Accuracy when the mistuning was in the violin line was .38 [.31, .46] and in the flute line was .20 [.16, .25].

**Discussion**

The interleaved melody paradigm was designed to examine whether musical training on a particular instrument increases timbral sensitivity to that instrument, using mistuning detection in real melodies rather than rhythm judgements for artificial tone sequences, as in Experiment 1a. Contrary to the hypothesis, results converge with Experiment 1a and 1b: musical training does not have an influence on timbre sensitivity, and support the alternate hypothesis proposed in Experiment 1a: attention influences perception. Similarly to the hypothetical orchestral line transcription described before, in this paradigm detection of mistunings, which first requires the separation of the melody from its context, did not depend on the instrument in which the mistuning appeared, but rather which line the listener's attention was directed to. The idea that attention influences perception is certainly not new (Carlyon, Cusack, Foxton, & Robertson, 2001; Dowling, 1990; Snyder, Gregg, Weintraub, & Alain, 2012; Spielmann, Schröger, Kotz, & Bendixen, 2014) but the above results suggest that attentional focus is more important than specific musical training in driving auditory stream segregation, leading to the lack of effect of specific instrumental musical training.

Table 3. Details of the mixed effects binomial logistic regression, where accuracy is predicted by fixed effects as described in the text and participant number as random effects on intercepts.

| Predictor | Estimate | p-value |
|---|---|---|
| Intercept | -2.43 | < .01 |
| Musicianship | 0.23 | .32 |
| Metrical mistuning location | 0.30 | .07 |
| Instrumental mistuning location | 0.47 | .03 |
| Top melody | 0.87 | < .01 |
| Attentional focus | -0.94 | < .01 |
| Pitch Overlap | 0.04 | .61 |
| **Random Intercepts** | **Variance** | |
| Participant | 0.02 | |

**General Discussion**

Though previous literature would suggest that instrumentalists are more sensitive to their own instrument's timbre (Margulis et al., 2009; Pantev et al., 2001), behavioural evidence for this claim was not found here. We instead propose that behaviour is guided by attention rather than musical training, consistent with literature exploring the effects of attention on auditory scene analysis (Andrews & Dowling, 1991; Bigand, McAdams, & Forêt, 2000; Jones, Alford, Bridges, Tremblay, & Macken, 1999; Macken, Tremblay, Houghton, Nicholls, & Jones, 2003). This interpretation was supported in both Experiments 1b and 2. In Experiment 1b, trials where rhythm captured attention more often also resulted in a segregated percept (from post-hoc analysis). In Experiment 2, attentional focus was a predictor of response accuracy for identifying mistunings, where accuracy depends on participants successfully streaming the relevant melody. Furthermore, performance when participants were asked to identify mistunings in both lines at once was particularly poor, highlighting the importance of attentional focus for successful task completion.

It is interesting to consider why the present results diverge from those found in cognitive-neuroscientific studies which have found instrument-specific effects of musical training. It may be that methods such as EEG, MEG and fMRI provide more sensitive measures that are capable of picking up on small effects of instrument-specific training which are not expressed in behavioural measures such as those used here. Greater sensitivity of neural over behavioural measures has been observed in research on processing dissonant and mistuned chords (Brattico et al., 2009) and harmonic intervals varying in dissonance (Schön et al., 2005). Alternatively, it may be that the instrument-specific effects observed in previous research were actually driven by greater attention to an instrumentalists' own instrument. Further research is required to disentangle these alternative accounts.

Let us now look at the ABA- paradigm more closely. Despite listeners most often initially perceiving maximally different timbres as segregated, there is still a fairly large proportion of trials heard as integrated. This was explained above by timbre similarity, but it may not be the only factor; based on personal listening, and participant feedback, the stimuli are clearly bistable, suggesting that timbre alone may not be enough to fully segregate two sounds played with same pitch, loudness and length. In a musical sense, this is very useful and is often employed by composers wanting to create instrumental chimerae or even simply writing passages involving the entire sections of the orchestra playing the same line. This suggests that timbre is a less important feature in perception of polyphonic music, with pitch, rhythm and loudness taking precedence. Relative importance of these four parameters for auditory streaming could be evaluated by combining parameters to see which causes streaming first. Some questions concerning salience and combining musical parameters in a streaming paradigm have been investigated (Dibben, 1999; Prince, Thompson, & Schmuckler, 2009; van Noorden, 1975) but a clear map of relationships between parameters has not yet been established, largely due to the complexity of polyphonic music. It might be a different situation for non-musical, or 'environmental' sounds, and both would be interesting to investigate further.

According to Horváth et al. (2001), predictive representations for both galloping and even patterns are held in parallel, but this was only tested where auditory stimuli were ignored. This explanation relies on predictive regularity to explain auditory scene analysis, as does the auditory event representation system model (Schröger et al., 2014). This model attempts to explain how auditory streams are formed right from the beginning rather than through a gradual buildup of evidence (Bregman, 1978), or through bistability (Pressnitzer, Suied, & Shamma, 2011). It builds chains of potential perceptual representations that compete for dominance; as new sounds are fed in, certain representations are validated and others are deleted until there is only one 'winner'. In terms of the ABA- paradigm, the model output is not conclusive as both percepts are valid and stable with respect to the model and so it does not help explain how auditory streams are formed in this paradigm. It seems attention is necessary to explain the creation of auditory streams when the stimuli could be interpreted in multiple ways, as is demonstrated in the experiments presented above.

To summarize, two streaming paradigms designed to investigate timbre sensitivity show that task performance depends not on sensitivity to a particular timbre due to instrument-specific musical training, but on allocation of attention to the appropriate, in the case of Experiment 2, or simply the chosen, in the case of Experiment 1a, auditory object.

**Acknowledgements**

Thank you to Lauren Stewart who provided access to MUMS and to Christoph Reuter and Andrew Staniland for help with constructing the morphing timbre stimuli.

# Appendix A – Stimuli experiment 1

| Timbre (original file name) | Pitch | Length (ms) | Peak Amplitude (dB) | Fadeout (ms) |
|---|---|---|---|---|
| Cello (CelA3_3.84sec) | A3 | | | |
| Cello (CelC#4_2.44sec) | C#4 | | | |
| Cello (CelD4_2.77sec) | D4 | 114 | -16 | 10 |
| Cello (CelE4_2.67sec) | E4 | | | |
| Cello (CelF4_2.56sec) | F4 | | | |
| Cello (CelF#4_2.12sec) | F#4 | | | |
| Trombone (TTbnG3_2.17sec) | G3 | | -12 | |
| Trombone (TTbnG#3_2.22sec) | G#3 | | | |
| Trombone (TTbnB3_2.54sec) | B3 | 113 | -15 | 10 |
| Trombone (TTbnD4_2.81sec) | D4 | | | |
| Trombone (TTbnD#4_3.54sec) | D#4 | | -16 | |
| Trombone (TTbnF4_3.01sec) | F4 | | | |
| Trumpet (CTptG#3_6.06sec) | G#3 | | -12.5 | |
| Trumpet (CTptA#3_2.75sec) | A#3 | | -16 | |
| Trumpet (CTptC4_7.44sec) | C4 | 111 | -13 | 10 |
| Trumpet (CTptD#4_3.54sec) | D#4 | | -12 | |
| Trumpet (CTptE4_7.42sec) | E4 | | -15 | |
| Trumpet (CTptF#4_6.55sec) | F#4 | | -14 | |

| | | | |
|---|---|---|---|
| Violin (VlnG3_8.79sec) | G3 | | -16 |
| Violin (VlnA3_8.98sec) | A3 | | -15 |
| Violin (VlnA#3_8.58sec) | A#3 | | -16 |
| | | 114 | | 10 |
| Violin (VlnB3_9.67sec) | B3 | | -12 |
| Violin (VlnC4_7.69sec) | C4 | | -14 |
| Violin (VlnC#4_7.12sec) | C#4 | | -16 |